\documentclass{IEEEtran}
\usepackage[utf8]{inputenc}

\usepackage{tabularx}

\usepackage[backend=biber,sortlocale=de,maxnames=20,style=numeric-comp,giveninits=true,isbn=false]{biblatex}
\begin{document}

\title{Reconstruction as a service: a data space for off-site image reconstruction in magnetic particle imaging
}
\author{Anselm von Gladiss, Amir Shayan Ahmadian, Jan Jürjens
\thanks{A. von Gladiss, A. S. Ahmadian and J. Jürjens are with the Institute for Software Technology, University of Koblenz, Koblenz, Germany (e-mail: vongladiss@uni-koblenz.de).}
}
	
\maketitle

\begin{abstract}
Magnetic particle imaging (MPI) is an emerging medical imaging modality which offers a unique combination of high temporal and spatial resolution, sensitivity and biocompatibility.
For system-matrix (SM) based image reconstruction in MPI, a huge amount of calibration data needs to be acquired prior to reconstruction in a time-consuming procedure.
Conventionally, the data is recorded on-site inside the scanning device, which significantly limits the time that the scanning device is available for patient care in a clinical setting.
Due to its size, handling the calibration data can be challenging.
To solve these issues of recording and handling the data, data spaces could be used, as it has been shown that the calibration data can be measured in dedicated devices off-site.
We propose a data space aimed at improving the efficiency of SM-based image reconstruction in MPI.
The data space consists of imaging facilities, calibration data providers and reconstruction experts.
Its specifications follow the reference architecture model of international data spaces (IDS).
Use-cases of image reconstruction in MPI are formulated.
The stakeholders and tasks are listed and mapped to the terminology of IDS.
The signal chain in MPI is analysed to identify a minimum information model which is used by the data space.
\end{abstract}

\section{Introduction}

Magnetic particle imaging (MPI) is an emerging medical imaging technique that images the spatial distribution of super-paramagnetic iron-oxide nanoparticles which serve as tracer material.
It is a functional imaging modality with tissue not being visible in the reconstructed images.
MPI was first introduced in 2005~\cite{Gleich2005} and relies on the magnetisation of the tracer material by magnetic excitation fields.
Due to the time-varying magnetisation, a voltage signal can be measured.
Introducing a magnetic gradient field enables spatial encoding and the reconstruction of the voltage signal into images showing the tracer's distribution.

MPI still holds pre-clinical status, but clinical use-cases are being identified and devices fit to clinical application are being constructed~\cite{Rahmer2017, Graeser2019, Vogel2023}.
Possible clinical applications range from diagnostics in e. g. oncology~\cite{Grafe2016, Mason2021} and angiography~\cite{Vogel2023} to theranostic approaches~\cite{Tay2018, Bakenecker2021}, interventional use-cases~\cite{Haegele2016, Wegner2020b}, visualising physical properties~\cite{Zhong2021}, stem-cell research~\cite{Zheng2016} and many more.
Key features of MPI are a spatial resolution in sub-millimetre range~\cite{VonGladiss2020}, a temporal resolution of more than 90 volumes per second~\cite{VonGladiss2020}, a current sensitivity limit of only 200~pg iron~\cite{Graeser2017c}, quantifiable measurements~\cite{VonGladiss2020} and bio-compatibility~\cite{Matschegewski2019}.

To reconstruct the measured voltage signals into images showing the distribution of tracer, mainly three techniques exist, which are system-matrix (SM) based reconstruction~\cite{Gleich2005, Knopp2010a}, x-space reconstruction~\cite{Goodwill2010} and neural-network based reconstruction~\cite{Hatsuda2016, VonGladiss2022a, Gungor2024}.
Currently, image reconstruction in MPI is performed on the same site, that is in the same facility, as the measurement is carried out.
Both now and in the future, when MPI is in a clinical state, an off-site image reconstruction may be beneficial, as medical doctors could access the latest image reconstruction algorithms and hospitals would not need to maintain image reconstruction servers.
Especially for SM based reconstruction, a huge amount of calibration data would not need to be handled on-site (see "Problem statement").

To this purpose, we propose a data space dedicated to medical image reconstruction.
The data space links measurement data of imaging facilities to corresponding off-site calibration data and offers image reconstruction as a service.

\subsection{Problem statement}

For an imaging facility such as a hospital to perform SM-based image reconstruction in MPI, there are two mayor challenges.
First, calibration data (the system matrix) needs to be measured in a time-consuming process. Second, the huge amount of calibration data needs to be handled efficiently for enabling fast image reconstruction.

There are mainly three image reconstruction methods in MPI, which are based on a system-matrix (SM), x-space and neural networks, respectively.
In this contribution, we will focus on SM-based reconstruction, as it seems particularly useful to combine with data spaces due to data handling.

SM based reconstruction has been introduced simultaneously with MPI itself in~\cite{Gleich2005} and is well-established especially when using multi-dimensional magnetic excitation fields.
Before reconstruction, the SM needs to be acquired.
Conventionally, this is performed in the imaging device itself.
A robot is used for sequentially positioning and measuring a sample of the tracer material at multiple spatial positions within the measurement field.
This set of calibrations measurements is stored as a SM encoding the system response to the tracer at different spatial positions.
Using the SM and the actual (patient) measurement, a linear system of equations can be set up and solved for the spatial distribution of the tracer.

The SM encodes the tracer material, measurement parameters such as the magnetic field, and also the hardware of the scanning device.
If only one of these characteristics changes, another SM needs to be acquired.
In~\cite{VonGladiss2021} it has been estimated that one clinical MPI scanning device would need to measure about four months to acquire 28 necessary SM.
As the scanning device is not available for patient care while calibrating, on-site calibration is not feasible in a clinical scenario.

After measuring the SM, it needs to be handled for image reconstruction.
One SM may have a memory demand of up to 117~GB.
For particular applications it may be necessary to use about 820~GB of memory for handling multiple SM at the same time according to~\cite{VonGladiss2021}.
Hence, dedicated image reconstruction hardware that is capable of fulfilling these memory demands is necessary.
Imaging devices are suitable for application in a mobile setting such as~\cite{Graeser2019} might be used in a bed-side device in a high-care unit or even in an ambulance.
Here, dedicated reconstruction hardware would contradict the demand of mobile devices.

In this work, these two challenges are addressed by proposing a data space enabling image reconstruction as a service.

\subsection{Related work}

It has been shown that a SM can be acquired in a dedicated device enabling off-site calibration.
This has been shown first for one-dimensional excitation fields~\cite{Gruettner2011}, but as well for multi-dimensional excitation fields~\cite{VonGladiss2017,VonGladiss2020a}.
Multiple corrections steps need to be applied in order to use an off-site SM for reconstructing images~\cite{VonGladiss2020a}.

SM may also be modeled off-site and can later be used for image reconstruction~\cite{Knopp2010c}.
However, the underlying particle model is usually a simplification of the complex particle behaviour and thus, reconstruction artifacts remain.

In order to exchange measurement data between multiple research groups, a data format called "Magnetic Particle Imaging Data Format" (MDF) has been developed~\cite{MDF}.
Beside the measurement signal itself, metadata may be stored.
As a research data format in a pre-clinical state, it is mostly patient-agnostic.
A subject name (if a subject would be scanned) must be provided in the current version of MDF, but may be replaced by an alias for the purpose of this work.

In the field of data spaces, this work relates to use-cases of data spaces in medical domain.
In medical data spaces, health data is integrated for different purposes, such as enlarging a data set for increasing the statistical analysis outcome or to improve or enable precision medicine~\cite{Berlage2022}.
Even the patient may be integrated in the data space by e. g. wearables~\cite{Raab2023}.
Often, medical data spaces aim at increasing patient safety.
One of the main issues identified is data governance and data privacy which must follow nation-specific legislation.

Although this work relates to medical data spaces, it differs from the mentioned publications as health data of different sources will not be integrated, but voltage signals and functional medical images that only reveal anatomical structures indirectly.
As this raw data can hardly be traced back to an individual patient, data privacy may be easy to guarantee in the proposed data space.
Further, it is not aimed for integrating the individual patient as a stakeholder, as usually the raw data of medical measurements are not comprehensible, and thus, need not be accessible to the patient.

International data spaces (IDS) have been founded as an initiative providing a cross-domain-data-space.
Within IDS, a reference architecture model (RAM)\footnote{https://docs.internationaldataspaces.org/ids-knowledgebase/v/ids-ram-4/} has been developed which will be used in this work.

\subsection{Structure}

In the next section, the required data for image reconstruction in MPI is described.
It is analysed which data needs to be shared in an off-site image reconstruction setting.
Use-cases for off-site image reconstruction are formulated.
Further, potential use-cases are identified that become feasible using data spaces.
Based on the identified data and use-cases, stakeholders are named and mapped to the terminology of IDS-RAM.
Thus, the applicability of IDS-RAM to the proposed data space is analysed.

\section{Minimum information model}

As mentioned above, correction steps need to be performed before using an off-site SM for image reconstruction.
These steps will be analysed especially as new demands may arise regarding data sharing.
A thorough review of the relevant data needed for image reconstruction is performed to identify which data need to be part of the proposed data space.

The measured signal in MPI can be expressed by
\begin{equation}
    u_k = -a_k \frac{\mu_0}{T}\int_\Omega\int_0^T \frac{\delta}{\delta t} \mathbf{M}(\mathbf{r},t)\cdot \mathbf{p}(\mathbf{r})e^{2\pi ikt/T}\textrm{d}t\textrm{d}^3r \, .
    \label{eq:mpi-v-signal}
\end{equation}
Here, $u_k$ is the $k$th frequency component of the measured voltage $u(t)$, $\mathbf{M}(\mathbf{r},t)$ denotes the time- and space-dependent magnetisation, $\mathbf{p}(\mathbf{r})$ is the sensitivity of the receive coil and $a_k$ is the $k$th frequency component of the transfer function of the receive chain, which consists usually of active filters, capacitors and analog-digital-converters.

The magnetisation $\mathbf{M}(\mathbf{r},t)$ depends on the tracer material (the amount and type of particles) and the applied magnetic fields and is the key component which needs to be measured, as simulations still lack precision, as mentioned above.
Hence, information about the type of tracer material and the applied magnetic fields must be shared for off-site image reconstruction.
The MDF already specifies that this information should be included when sharing data, although information about the tracer material is still marked as "optional", which should change when applying our use-case.

The coil sensitivity $\mathbf{p}(\mathbf{r})$ is device-specific.
In a scanning device featuring small receive coils such as in spectrometric devices, the sensitivity is often assumed to be homogeneous such as in~\cite{VonGladiss2020a}.
With increasing measurement field size, an inhomogeneous coil sensitivity becomes more probable.
Thus, it can be assumed that the coil sensitivity needs to be corrected when using an off-site SM for image reconstruction.
The coil sensitivity does usually not change for the lifetime of one receive coil and can be initially measured when integrating the coil into the measurement setup.
The sensitivity profile can be stored on-site (e. g. with the hospital) or off-site (e. g. with the manufacturer) but needs to be applied to the off-site SM before reconstruction.
If being stored on-site, it may be integrated into the MDF first using a "user-defined parameter" and second by an updated version of the MDF.

The transfer function $\mathbf{a}$ characterises the receive chain of a measurement device and is specific as well.
Correcting a measurement for the transfer function is always necessary when comparing measurements coming from different devices and is crucial for image reconstruction using an off-site SM.
The transfer function can be estimated~\cite{Knopp2010c} or measured directly~\cite{VonGladiss2020a}.
It remains constant as long as the receive chain is not modified or damaged.
It may be stored both on-site and off-site and needs to be applied to the measurement before reconstruction.
If it is not applied directly after the measurement, it can also be included in the data that is transferred and is already specified within the MDF.

Additionally to the parameters presented in~(\ref{eq:mpi-v-signal}), an empty measurement is usually performed and subtracted from the actual measurement in order to cope with background signal coming from the measurement device.
Same as the transfer function, it may be applied on-site or integrated using the MDF.
The latter might be beneficial as then reconstruction algorithms such as presented in~\cite{VonGladiss2017} may be used.

To summarize, the information that is needed for reconstructing an image off-site are:
\begin{itemize}
    \item the measurement / voltage signal in full representation or a selection of relevant frequency components as described in~\cite{Knopp2010a}
    \item information about the tracer material
    \item metadata about the applied magnetic fields
    \item the sensitivity profiles of the receive coils
    \item the transfer function of the receive chain, if it has not been applied to the measured signal
    \item an empty measurement, if one has not yet been subtracted from the measured signal
\end{itemize}
This information can be stored in a file stored in the MDF.

\section{Use-case analysis}

In this section, two potential use-cases for off-site image reconstruction are elaborated, which are online reconstruction of a live measurement and second-look image reconstruction with a most recent reconstruction algorithm.
Further, two scenarios are described that are potentially enabled using data spaces, namely the long-term measurement analysis for maintenance purposes and the collection of reconstruction data for training of machine learning algorithms.

The use-cases will be presented and analysed regarding their feasibility to be realised by the proposed data space.

\subsection{Online reconstruction of a live measurement}

In an interventional setting in a hospital, a coronary stent is positioned using MPI.
As measurement parameters, a 3D magnetic field is applied.
Both the blood flow and the stent are to be visualised using multi-color MPI~\cite{Rahmer2015}.
Therefore, two particle types are being used, one as part of the stent coating and the other as an injected particle bolus.
Inside the hospital, the voltage signal over three receive coil pairs is recorded and an empty measurement that has been recorded prior is subtracted.
As the hospital lacks both calibration data of the scanning device and a dedicated reconstruction server for MPI data, an MDF-file is created automatically holding information about the magnetic field and the two tracer materials, which is extracted from the scanning device and the radiological imaging system.
The MDF file is enriched by the coil sensitivity profile and the transfer function of the receive chains which are stored with the scanning device.

Via a data space connector, the hospital chooses the manufacturer of the scanning device to provide calibration data and to reconstruct the images.
The broker of the data space establishes a connection between hospital and manufacturer.
The manufacturer has a dedicated device to perform the calibration data measurements and therefore, correction steps need to be applied to the signals before reconstruction.
The coil sensitivity profiles of the receive coils is already stored at the manufacturer, so the profiles stored in the MDF file are not needed.
The coil sensitivity profiles of the imaging device at the hospital are applied to the off-site SM with the manufacturer.
Then, the transfer function of the receive coils is read from the MDF file and is applied to the voltage signals.
As multiple sequential measurements are to be reconstructed online, a data stream or a file stream between hospital and manufacturer is established.
Then, an iterative Kaczmarz algorithm is started at the image reconstruction server with the manufacturer.
The reconstruction parameters are determined automatically such as presented in~\cite{Scheffler2023}.

The reconstructed multi-channel images are send back to the hospital as a data stream or file stream in MDF file format.

As the clearing house logs the reconstruction request by the hospital, it may be invoiced instantly or on a regular basis.

\paragraph{Online reconstruction}

In a several potential clinical application of MPI the availability of online reconstruction is crucial, especially in interventional settings such as presented here (positioning a stent).
Thus, it must be ensured that the off-site character of image reconstruction does not become the bottleneck, but still enables online reconstruction.
The transfer of measurement data must not inhibit the live display of reconstructed images.

In MPI, data is sampled in a low MHz-range, e. g. 1.25~MHz.
When using three receive coil pairs in a 3D scanning device and assuming that the digital-to-analog converters feature quantisation levels of maxmimum 32bit, 15~MByte of data is generated each second without using signal compression techniques.

Nowadays, internet connections based on fibre optics feature upload data rates of 25~MByte per second for a private household and outreach the data rates in MPI.
Thus, off-site online reconstruction is feasible.

\begin{table*}[t]
    \centering
    \footnotesize
    \begin{tabular}{m{0.3\textwidth}|m{0.3\textwidth}|m{0.3\textwidth}}
    \textbf{stakeholder} & \textbf{possible roles} & \textbf{responsibilities} \\\hline
    
        imaging facility: hospital, radiological department &
        data customer, service consumer, app owner &
        initiator of transactions \\\hline
        
        calibration facility: manufacturer, tracer vendor, dedicated calibration institution &
        data owner, service provider, app owner &
        provides calibration data \\\hline
        
        reconstruction expert: manufacturer, dedicated reconstruction institution &
        app owner, service provider, data consumer &
        provides image reconstruction and signal analysis techniques \\
    \end{tabular}
    \caption{Overview about the roles in the proposed data space according to the business layer of IDS-RAM version 4 for image reconstruction use-cases.}
    \label{tab:stakeholder}
\end{table*}

\subsection{Experimental reconstruction with a most recent reconstruction algorithm}

In this scenario, an already completed measurement is reconstructed for a second time, as the reconstruction algorithm used priorly has generated reconstruction artifacts.
As in the previous scenario, an MDF file containing the measurement is generated and enriched with metadata.
The radiologist at the hospital chooses the appropriate sophisticated reconstruction algorithm and calibration data using the data space connector.

Again, the broker establishes the connections, guarantees for the data transfer and logs the transaction.
The hospital receives back the reconstructed image.

\subsection{Long-term measurement analysis for maintenance purposes}

In MPI scanning device, signal chains change over time or become unstable due to aging solder connections, broken capacitors or distorted filters.
In the send chain, this may result in varying current amplitudes and phases that need to be applied for generating the desired electromagnetic fields.
In the receive chain, these changes are visible in the receive signal, especially in the background signal obtained by an empty measurement.

Currently, data regarding the send chain is stored on-site and can be reviewed for maintaining purposes.
With an established data space, this data and also empty measurements could be transferred to the manufacturer on a regular basis.
Then, the manufacturer is able to track the signals' development automatically and prevent temporal failure of a device by on-time-maintenance.

\subsection{Collection of reconstruction data for training of machine learning algorithms}

Another reconstruction approach than using a SM is using neural networks.
In order to train a (deep) neural network, an appropriate amount of training data is needed.
Currently, the neural networks used to reconstruct MPI data are trained with simulated data.
As the simulated data does not incorporate the correct particle behaviour, the transferability to actual measurement data is questionable.

For other imaging modalities, it usually is a great effort to collect training data.
In the proposed data space, the collection of training data may happen incidentally.
If the imaging facility, that is the hospital, agrees to further usage of the measured data, it can be stored together with the reconstructed images by one participant of the data space.
This applies to both single images but also to time-series data.

Note that as a pure functional imaging modality and a patient-agnostic file format, data privacy will not be violated.

\section{Analysis of data space roles}

In the previous sections, relevant data for off-site image reconstruction has been identified and potential use-cases for a data space have been described.
The proposed data space shall be designed according to the reference architecture model (RAM, version 4) of international data spaces (IDS).
In this section, the IDS objects, the stakeholders and their roles according to the business layer of IDS-RAM will be identified within the described use cases.
Despite their importance to a data space, the roles of the broker and the clearing house will not be elaborated here, as they do not actively contribute to image reconstruction.

\subsection{IDS objects}
The data in terms of an IDS object is mainly the calibration data, even though it need not be transferred.
Within this work, it has been mentioned that MDF format is suitable for storing MPI data.
Thus, the MDF specification may be used as a vocabulary within the data space.
The image reconstruction process, which is the application of a reconstruction algorithm, is referred to as a service.
The image reconstruction algorithm itself will be called an app.

\subsection{IDS roles}

An overview about the stakeholders relevant to image-reconstruction-scenarios is given in Table~\ref{tab:stakeholder}.
Three essentials roles concerning off-site image reconstruction have been identified in the data space, which are imaging and calibration facilities and the reconstruction experts.
Further, the clearing house and the broker are crucial infrastructural roles that are naturally part of the data space.

The imaging facility usually fills in the role of the data customer.
However, as off-site image reconstruction is performed, it will not be data consumer, but service consumer.
In case the imaging facility owns the reconstruction algorithm (the app) and provides it to another facility for actually performing the reconstruction, it is also an app provider.

The calibration facility owns the calibration data.
It may also own the reconstruction algorithm and provide the service of image reconstruction.
Further, it may provide other services such as signal analysis for maintenance purposes as described above.
In the latter case, the imaging facility would fill in the role of data owner as well.

A dedicated image reconstruction facility typically owns the reconstruction algorithm (app owner) and provides the service of image reconstruction.
Then, it fills in the role of data consumer, as it needs the calibration data of the calibration facility.
It may also provide the app to e. g. the calibration facility, if image reconstruction will be executed there.

While the role of imaging facilities is usually filled in by hospitals or radiological departments, the other roles may be filled in by dedicated institutions or the manufacturer of the scanning device.
In the latter case, the data space may shrink to a connection between the devices manufacturer and the imaging facility (beside broker and clearing house).

However, the different responsibilities of the roles allow for a broad network of specialists in their field.
Dedicated institutions for calibration measurements may provide SM for different scanning device topologies and may have expert knowledge about compression methods that speed up reconstruction algorithms~\cite{Knopp2015a}.
Patient safety may be enhanced by integrating reconstruction experts in the daily clinical image reconstruction.
Then, the latest and approved image reconstruction algorithms may be used for reducing image artifacts and improving diagnosis.

\section{Conclusion and discussion}

\subsection{Principal Results}

A data space has been proposed that offers off-site image reconstruction with off-site calibration data.
This is advantageous for an imaging facility, as it does not need to acquire calibration data in a time-consuming measurement.
Further, it does not need dedicated hardware to handle huge matrices or to enable fast image reconstruction.
Instead, the imaging facility may choose to use the latest reconstruction algorithms on the market.

For the vendor/manufacturer of the scanning device, who usually sells this hardware and the corresponding algorithms, another business model is created.
Switching from the one-time-sell of hardware and algorithm to a pay-what-you-need-model, another steady source of income is generated beside the maintenance contract.
Beside the feasibility of off-site online image reconstruction, the proposed data space can also be used for optimising maintenance cycles of the imaging devices by the manufacturer.

A minimum information model has been identified for sharing the measurement data and reconstructed images.
The MDF is an appropriate file format to store the minimum information.

\subsection{Limitations}

This concept for a new medical data space has not been implemented and tested so far.
In future work, a reference implementation will be provided and research institutions will be invited to join the data space, together with the manufacturers of pre-clinical MPI scanning devices.

As MPI has not been approved yet for human use, hospitals will currently not connect if not for research purposes.
Nevertheless, this is the right time to create the infrastructure to enable off-site image reconstruction for MPI.
As soon as MPI has been approved for clinical use, the data space can be scaled accordingly.

\subsection{Conclusions}

A data space for image reconstruction in MPI has been proposed.
Basically, the principles of this data space apply to other medical imaging modalities as well.
Also in other imaging modalities such as computed tomography or magnetic resonance imaging hospitals might benefit from the availability of a broad range of image reconstruction algorithms.
This holds as well for X-space based MPI reconstruction, an alternative to SM based reconstruction.
However, SM based reconstruction in MPI seems especially suitable for this data space, as it enables to use off-site calibration data.

\printbibliography

\end{document}